\documentclass{article}

\usepackage{graphicx}
\usepackage{rotating} 
\usepackage{fullpage,amsmath,amssymb,latexsym}
\usepackage{multirow}
\usepackage{natbib}

\begin{document}


\title{Metallicity of the Massive Protoplanets Around HR 8799 If Formed by Gravitational Instability}


\author{R. Helled$^1$ and P. Bodenheimer$^2$\\
$^1$Department of Earth and Space Sciences,\\
University of California, Los Angeles, CA 90095 1567, USA\\
rhelled@ucla.edu\\
$^2$Department of Astronomy and Astrophysics,\\
UCO/ Lick Observatory, University of California, Santa Cruz, \\
CA 95064, USA\\
peter@ucolick.org}

\date{}
\maketitle 

\begin{abstract}
The final composition of giant planets formed as a result of gravitational
instability in the disk gas depends on their ability to capture solid material
(planetesimals) during their 'pre-collapse' stage, when they are extended and
cold, and contracting quasi-statically. The duration of the pre-collapse stage is
inversely proportional  roughly to the square of the  planetary mass, 
so massive protoplanets have
shorter pre-collapse timescales and therefore limited opportunity for planetesimal
capture. The available accretion time for protoplanets with masses of 3, 5, 7, and
 10 Jupiter masses is found to be 7.82$\times 10^4$,  2.62$\times 10^4$,
 1.17$\times 10^4$ and  5.67$\times 10^3$ years, respectively.
The total mass that can be captured by the protoplanets depends on the planetary
mass, planetesimal size, the radial distance of the protoplanet from the parent
 star, and the local solid surface density. 
We consider three radial distances, 24, 38, and 68 AU, similar to the radial
 distances of the planets in the system HR 8799, and estimate the mass of heavy
elements that can be accreted. We find that for the planetary masses usually
adopted for the HR 8799 system, 
the amount of  heavy elements accreted by the planets is small, leaving them
 with nearly stellar compositions.  
\end{abstract}

\newpage

\section{Introduction}
 The recent discoveries of wide-orbit massive protoplanets by
 direct imaging (Kalas et al., 2008; Marois et al., 2008) have presented new type of gaseous planets
 that giant planet formation theories must be able to explain. Work regarding possible formation scenarios has already been 
 demonstrated by several groups (e.g., Dodson-Robinson et al., 2009b; Nero and Bjorkman, 2009). Systems with massive gaseous planets at large radial distances offer a challenge for theorists mainly because they cannot be explained easily by the core accretion model, the standard model for giant planet formation (e.g., Cameron, 1978; Pollack et al., 1996). The core accretion model fails to form giant planets at large radial distances in situ due to the low surface density and the extremely long accretion timescale (Ida and Lin, 2004; Dodson-Robinson et al., 2009b). As a result, a more detailed investigation of giant planet formation scenarios seems to be required. 
 \par
 
 In addition, the increasing number of transiting planets provides information regarding the
 planetary mean densities, and therefore, their compositions. Giant planet
 formation theories are also required to explain the planets' observed properties, and when 
possible, provide predictions that can be tested with future observations and
 upcoming data. Since the heavy element masses of planets can be estimated for
 transiting planets, further investigation and more detailed understanding of
 the origin of the heavy elements and the cause for the large variety in heavy
 elements enrichments observed (e.g., Guillot et al., 2007) is desirable. \par

While the core accretion model offers an enrichment with heavy elements that is
 coupled to the formation process (e.g., Pollack et al., 1996; Hubickyj et al.,
 2005), in the gravitational instability model (Cameron, 1978; Boss, 1997) the
 planet's enrichment with solids is so far considered as an independent part of 
the actual formation mechanism. 
In the disk instability model giant protoplanets are initially formed with stellar
 compositions, and can be enriched with heavy elements after their formation by
 accretion of planetesimals (e.g., Helled et al., 2006). The total mass of solids
 that can be accreted depends on the planetary mass, its radial distance, the
 available solid material for capture, and the protoplanet's efficiency in 
accreting the solid planetesimals (Helled and Schubert, 2009).
\par

A gravitationally unstable condensation of a few Jupiter masses in a
protoplanetary disk evolves through three phases (DeCampli and Cameron, 1979;
Bodenheimer et al., 1980). First, it contracts quasi-statically with cool
internal temperatures on the order of a few hundred K, with hydrogen in
molecular form,  and with radii a few thousand times the present Jupiter radius.
Second, once the central temperature reaches about 2000 K, the H$_2$
dissociates and initiates a dynamical collapse of the entire planet, ending
only when the radius has decreased to a few times the present Jupiter radius.
Third, it contracts and cools on the long time scale of $\sim 10^9$ yr.
The first phase is the crucial one for the   capture of solids, and the time
scale of this phase, which is about $4 \times 10^5$ yr for 1 Jupiter mass,
determines the amount of material that can be captured. 
 As the protoplanets contract quasi-statically during this phase, planetesimals
 in their feeding zone can be slowed down by gas drag, and get absorbed in the
 planetary envelopes (e.g., Helled and Schubert, 2009). 
As a protoplanet contracts, fewer planetesimals pass through its envelope, and 
instead of being accreted they get ejected from the protoplanet's vicinity.
 Efficient planetesimal accretion ends once the central temperature reaches
 $\sim$ 2000 K and  molecular hydrogen begins to dissociate.

Whether gaseous protoplanets can form as a result of a local gravitational
 instability in a protoplanetary disk is still a matter of debate. The majority of
 theoretical and numerical work suggests that the gravitational instability
 scenario for planet formation is rather unlikely, at least at relatively small
 radial distances (e.g., Rafikov, 2007; Cai et al., 2009). Other results suggest
 that protoplanetary disks can break into gaseous protoplanets (e.g., Boss, 1997), although this may be limited to special
conditions (Mayer et al., 2007). The extrasolar giant planets observed by direct imaging (Kalas et al., 2008; Marois et al., 2008) have raised the
 suggestions that such massive gaseous planets at large radial distances were
 formed as a result of gravitational instabilities (Dodson-Robinson et al., 2009b;
 Nero and Bjorkman, 2009). The fact that HR 8799 host star is found to be metal-poor ([Fe/H]=-0.47) supports the idea that its planetary system was formed via disk instabilities. 
More advanced numerical simulations with longer integration times and further
 theoretical work seem to be required before a robust conclusion regarding this
 formation mechanism and its limitations can be made.  While recent theoretical
work (Boley, 2009; Rafikov, 2009; Nero and Bjorkman, 2009) suggests that gravitational instability can
form planets only at distances greater than 50-100 AU, it is possible that these
planets can migrate inward as a result of interactions with the disk, after
formation. Thus  
in this paper we use the working hypothesis that massive gaseous planets presently
at radial distances $>$ 20 AU could have formed  as a result of gravitational instability. 
Below we investigate the process of heavy element enrichment via planetesimal
 capture for the planetary system HR 8799. 

\section{Enriching the Protoplanets with Solids}

\subsection{Pre-Collapse Evolution}

Even under the assumption that gaseous protoplanets can form as a result of
 gravitational instability in the disk, a complication comes from the fact that
 the initial states (configurations) of the formed protoplanets are unknown.
 Numerical simulations of protoplanetary disks often report different
 configurations for the newly formed gaseous objects. Simulations made by Boss
 (e.g., Boss, 1997, 2002) typically result in colder and less dense objects than
 the ones found by Mayer and collaborators (Mayer et al., 2002, 2004, 2007).
Both of these groups present denser and warmer initial configurations than the
 ones used in previous investigations (DeCampli and Cameron, 1979; Bodenheimer
 et al., 1980). \par

We take the protoplanets to be  spherical and static objects.  The initial
radius must be less than the Hill radius. We assume
 that the planets  were formed at their observed locations namely, 24, 38, and 
68 AU, or at least reached those positions early in their evolution.  The
 planetary masses for the HR 8799 system are estimated to fall in the range 
5-13 Jupiter masses (Marois et al., 2008); we consider planets with masses between
3 and 10 Jupiter masses. We take the initial configuration such that the
 protoplanet is gravitationally bound at a radial distance of 24 AU 
 around a 1.5 M$_{\odot}$ star, the minimal radial distance of the HR 8799 system.
  
Typically, the chosen initial states of protoplanets can be taken as arbitrary,
 based on the argument that after a relatively short time the planets retain no
 memory of their initial states (DeCampli and Cameron, 1979; Marley et al., 2007;
 Helled and Schubert, 2009). However, if the evolutionary time is short, as in the
 case for massive gaseous protoplanets, different initial configurations can 
result in different pre-collapse timescales. While the different initial states do 
not affect much the global evolution of the planets, they may influence the
 accretion process. This issue is further discussed in section 3.\par 

Our baseline case for the initial model is a 'cold start' in which the
 protoplanets are as extended and cold as possible (and yet gravitationally bound),
 resulting in the maximal pre-collapse contraction time, and therefore the longest
 available time for planetesimal accretion. We then follow the planetary evolution
 up to the point when the central temperature reaches about 2000 K and the 
dynamical collapse begins. The four standard equations of stellar structure are
solved (Henyey et al., 1964): hydrostatic equilibrium, mass distribution, energy
conservation,
and radiative or convective transport.  The energy source is gravitational
contraction of the gas alone. The surface boundary condition is that of
an isolated stellar photosphere. Calculated surface temperatures turn out
to be a few tens of degrees K, comparable to disk temperatures at the 
distances considered; thus disk irradiation effects will not have  
significant influence on the boundary condition. In addition, as we show below, efficient planetesimal accretion occurs when the protoplanets have already evolved to denser and warmer configurations with the surface temperatures being higher than the disk temperatures. As a result, the  disk temperature becomes insignificant by the time planetesimals can be accreted. No central solid core is present. 
The low-temperature opacities, which are dominated by grains with an
interstellar size distribution, are obtained from Pollack et al. (1985) and
Alexander and Ferguson (1994).   The equation of state is based on the tables
of Saumon et al. (1995).
The  accreted mass of solid planetesimals is computed during the pre-collapse evolution. \par

Table 1  summarizes the initial states for planetary masses between 3 and 10 M$_J$
 (M$_J$ being Jupiter's mass).  The last column provides the precollapse time-scale
 for each mass. 
Figure 1 presents the evolution of protoplanets with masses of 3, 5, 7, and 10
 Jupiter masses. As can be seen from the figure, more massive protoplanets have
 shorter pre-collapse time-scales. The initial radius increases  with increasing
 planetary mass, as more massive objects can be more extended and still be
 gravitationally bound due to stronger gravity. The central temperature however,
 is higher for more massive bodies. \par
 
 It is possible that massive clumps ($\sim$ 10 M$_J$) do not form directly by a single gravitational 
 instability, but are formed as a result of substantial gas accretion and possibly mergers of smaller clumps. In such a case the effective pre-collapse contraction timescales of the massive protoplanets would be longer than presented here and comparable to the timescales which correspond to smaller objects. As a result, the final  metallicity of the massive protoplanets would be determined by the planetesimal accretion efficiency of the smaller protoplanets. However, as we show below, for the range of masses we consider, there is no significant difference in the resulting metallicity of the protoplanets. \par
 
Our evolutionary model assumes a non-rotating clump, while in
reality giant gaseous protoplanets formed by gravitational instabilities are likely to have some angular momentum.
Most of the interior of the protoplanet is convective, and its effect will be to transport angular momentum outward, on
a convective turnover time scale, and drive the distribution of angular momentum toward uniform
rotation. The resulting structure, which still needs to be investigated in detail, will have most of the mass in slow
rotation, and a relatively small fraction of the mass in an outer subdisk. The effective mass of the protoplanet, which
represents the low-angular momentum portion, will be somewhat lower than the total mass we consider and  will therefore have
 a somewhat longer contraction time. Also, the formation of the subdisk can result in a longer available time for solid accretion since it is likely to exist after the dynamical collapse, providing a source of gas drag that can lead to additional accretion. In addition, planetesimals in such a disk can collide and lead to the formation of satellites (Helled et al.,~2006). 

\subsection{Planetesimal Accretion}

The amount of solid material (heavy elements) accreted by a gaseous protoplanet depends on the available mass for capture in the planet's feeding zone, and the protoplanet's efficiency in accreting the solid planetesimals. 
The planetesimal accretion rate is given by (Safronov, 1969)
\begin{equation}
\frac{dm}{dt}=\pi R^2_{capture} \sigma \Omega F_g,
\end{equation}
where $R_{capture}$ is the protoplanet's capture radius (see Helled et al., 2006
 for details), $\sigma $ is the solid surface density, and $\Omega $ is the
 protoplanet's orbital frequency. We take the gravitational enhancement factor $F_g$
 (Greenzweig and Lissauer, 1990) as unity, providing a lower bound estimate for the capture rate. 
It should be noted, that Safronov's formulation (equation 1) assumes that the
 protoplanet is rotating around the star with the same angular velocity as the
 gas.  The protoplanet's motion relative to the gas is in the $z$-direction, due
to its inclined orbit. The formulation applies when assuming that the planetary 
orbit goes from the "bottom" of the disk, through the midplane and to the "top" of 
the disk in half an orbit, so in every orbit the protoplanet goes the entire
 thickness of the disk encountering solid surface density $\sigma$ (g cm$^{-2}$).
\par
We follow the evolution of the protoplanet, and at each stage we
calculate planetesimal trajectories with increasing impact
parameters, accounting for gas drag and gravitational forces,
to find $R_{capture}$. Ablation and fragmentation
of planetesimals are included in the trajectory calculation. We determine the largest impact parameter for which
planetesimal capture is possible. For this trajectory the planetesimal's closest distance of
approach to the protoplanet's center is defined as the
capture radius, $R_{capture}$ (see further details in Helled et al., 2006, Helled and Schubert, 2009 and references therein).  
It should be noted that the evolution calculation does not account for the accreted
material. However, as we show below, in all the cases considered in this work, the
 amount of solids being captured is negligible compared to the planetary mass and 
therefore is unlikely to affect the planetary evolution. In addition, we find that the 
solid material is deposited in the denser inner regions which are convective, and therefore the accreted material is not expected to change the opacity near the planet's  photosphere. As a result, the larger concentration of heavy elements would
increase the opacity but not affect the convective stability, so that the expected thermal evolution would be similar. More details
on the effect of planetesimal accretion on the thermal evolution, including the effect on the energy budget of the protoplanet, can be
found in Helled et al. (2006). \par

At early stages of the planetary evolution, the capture radius can be substantially
smaller than the physical radius of the protoplanet due to the low density in the
upper envelope, and the fact that the planetesimal must pass through denser inner 
regions to encounter enough gas drag, lose kinetic energy, and get captured by the 
protoplanet.  
Recently, major progress in understanding planetesimal formation has been achieved.
Although the process that leads to planetesimal formation is still debated, there
seems to be a general agreement that planetesimals have larger initial sizes than 
previously thought. Two independent groups have shown recently that large
planetesimals, of about 100 km in size, are formed directly from solid particle
concentrations, therefore 'skipping' sizes of hundred meters and a few kilometers
(e.g., Johansen et al., 2007; Cuzzi et al., 2008). Further  support for this idea 
came from yet another group that showed that planetesimals must form in big sizes
 in order to fulfill the constraints coming from the asteroid belt size-frequency
distribution (Morbidelli et al., 2009). We consider planetesimals with sizes of 1, 10, and 100 km. 
The planetesimals are
taken to be composed of a mixture of ice, silicates and CHON with an average
density of 2.8 g cm$^{-3}$ (see Pollack et al., 1996, and Helled et al., 2006
 for details). Although planetesimals are preferentially captured toward the
center of the protoplanet, their settling toward the center to form a core
is not considered. \par

As shown below, the accreted mass is rather sensitive to the assumed planetesimals' sizes. 
Planetesimals with sizes larger than 100 km will be even harder to accrete, resulting in lower captured mass than the results presented. Smaller planetesimals than the ones considered in this work will be captured more easily, and would lead to larger enrichment with heavy elements. Practically, the disk is likely to have a size distribution for the planetesimals, which can change with time (collisions, breakup) and radial distance. 
For simplicity, and since the real sizes of planetesimals and their time/radial dependence are not well constrained, we consider only three planetesimal sizes. 
\par

Naturally, the large orbital distances in the HR 8799 planetary system (Marois et
 al., 2008) lead to smaller enrichment due to the decrease in the solid surface
 density with radial distance, and slower accretion rate due to the dependence on
 the orbital frequency (see equation (1) and Helled and Schubert, 2009). 
We adopt the disk model presented by Dodson-Robinson et al. (2009b); this stellar-heated disk has
a Toomre Q value $\sim$ 1.5 around a 1.5 M$_{\odot}$ star. The surface density of the disk goes $\propto a^{-12/7}$ with $a$ being the radial distance from the star (see their eq. 6); the disk mass is about 30\% of the system mass. 
The solid/gas ratio is taken from the disk model of Dodson-Robinson et al. (2009a), which takes into account the details of the
chemistry at each distance. The distribution of the solid surface density is plotted in Figure 1 of Dodson-Robinson et al. (2009b). The solid surface densities at 24, 38, and 68 AU are found to be 7, 3, and 1
 g cm$^{-2}$, respectively. \par 
  
The planetesimal velocities far from the protoplanet are taken with respect to the protoplanet, i.e., 
the protoplanet is assumed to be at rest relative to the planetesimals, with the velocities taken to be 10\% of the
 Keplerian velocities, in agreement with estimates based on detailed 3-body 
interactions (e.g., Greenzweig and Lissauer, 1990). The velocities are therefore set to
 7.44$\times 10^4$, 5.92$\times 10^4$, and 4.42$\times 10^4$ cm s$^{-1}$, for 24,
 38, and, 68 AU, respectively. Once the planetesimal velocities for the different radial distances are set, the protoplanets' cross section for planetesimal capture, and the mass accretion rate can be computed. Gravitational
potential fluctuations in the disk as a result of gravitational instability could in fact stir up the planetesimal velocities
to higher values. The result would be a reduction in the planet's cross section for capture. 

Figure 2 shows the capture radii for 3 and 10 Jupiter mass protoplanets as a
function of time. The black, blue, and red curves correspond to radial distances
of 24, 38, and 68 AU, respectively. The dotted, dashed, and dashed-dotted curves 
represent 1, 10, and 100 km-sized planetesimals, respectively. As expected, larger
planetesimals result in smaller capture radii. 100 km-sized planetesimals cannot
be captured at very early stages of the planetary evolution due to low densities 
and temperatures at the upper envelopes of the protoplanets. A more detailed
 analysis on the behavior of the capture radius with time can be found in Helled
 et al. (2006). The capture radius increases with increasing radial distance
 due to lower planetesimal velocities. However, as we show below, due to the slow
accretion rate farther from the star, the mass that can actually be captured
 decreases with radial distance. 

\section{Results}

Table 2 presents the captured planetesimal mass for all the cases considered.  Larger planetesimals are harder to capture, and for all cases considered in this
 work the captured mass from 100 km-sized planetesimals is smaller than
 1 M$_{\oplus}$. If planetesimals are indeed formed large, massive protoplanets
 could not be enriched with heavy elements via planetesimal capture. 
A 10 M$_J$ protoplanet is massive enough to have a large capture capture cross
section, and therefore has a potential to capture a significant amount of solids.
However, it has a very short pre-collapse stage and therefore, the final captured
 mass is substantially smaller than the available mass for accretion. Nevertheless,
 a 10 M$_J$ planet at 24 AU can capture about 90 M$_{\oplus}$ of heavy elements 
if planetesimals are 1 km in size. 
As the planetary mass decreases the gravitational pull decreases but the available
time for planetesimal capture increases. As a result, a 3 M$_J$ protoplanet can 
accrete more solid material than  one of  5 M$_J$.

To check the robustness of our result and its sensitivity to the initial model of
 the protoplanet, we run a test case in which we allow a more compact and hotter
 initial model for a 5 Jupiter mass object. 
The properties of the initial model are listed in Table 1 (5 M$_J$ - hot start).
 The accreted planetesimal mass for this case is given in Table 3. As can be seen
 from the table, the numbers do not change substantially and the trend found in
the cold start case is certainly maintained. Although the pre-collapse timescale 
is shorter for protoplanets with a 'hot start', the accreted mass is nearly the same as for the 
cold start case. This is due to the fact that at the early stages of the cold start evolution the densities are 
low and only a very small fraction of planetesimals can be accreted before the hot start densities are reached.  
While the exact values of the accreted mass can change for different chosen
 initial conditions, the conclusion that massive protoplanets cannot accrete a
 significant amount of heavy elements at wide orbits, and that the mass decreases
 substantially with increasing planetesimal size is robust with regard to the assumed initial
condition for the protoplanet.  

\section{Discussion and Conclusions}

We investigate the possibility of heavy element enrichment for the planetary
 system HR 8799. We consider planetary masses of 3, 5, 7 and 10 M$_J$, with radial
 distance of 24, 38, and 68 AU. 
We find that planetesimal accretion is relatively inefficient for such a system
 due to main two reasons: First, massive protoplanets have relatively short
 precollapse stages, and therefore, have limited time for planetesimal capture.
 Second, the accretion rate is low at large radial distances due to its dependence
 on the solid surface density and orbital frequency (both decreasing with radial
 distance). Farther from the star planetesimal velocities are lower, but the
 accretion time-scale is substantially longer. \par

Smaller planetesimals can be accreted more efficiently, and if planetesimal sizes
 are of the order of $\sim$ 1 km, solid material between a few and tens 
of M$_{\oplus}$ can be captured. However, if planetesimals are formed with
 initial sizes of $\sim$100 km or larger, the accreted mass is negligible. We 
therefore conclude that under such conditions wide-orbit gaseous planets,
 if formed via gravitational instability,  will have nearly stellar compositions.
 \par

The capture cross-section used in this work was calculated taking the gravitational
 enhancement factor $F_g$ which corrects for the three-body effects
 (see Greenzweig and Lissauer 1990 for details) as unity. As a result, the accreted 
mass presented here should be taken as a lower bound. The total mass of solids in
 the protoplanet's feeding zone, which represents the maximum solid mass available
 for accretion, can be up to 2-3 M$_J$,  increasing with radial distance. However,
 it is unlikely that this available mass can actually be accreted due to the
 low accretion rate (see Helled and Schubert, 2009 for details) unless it is found
 that a significant gravitational focusing occurs. Detailed calculations of the
 gravitational enhancement factor for extended massive protoplanets with gas drag
 included seem to be required before a robust estimate for the upper bound of the
 accreted mass can be made.  
\par 

Our results suggest that massive protoplanets at wide orbits will have metallicity similar to their parent star. This conclusion does not depend on the stellar metallicity. Naturally, protoplanets around metal-rich stars will contain more high-Z material in their interiors due to the larger fraction of heavy elements (higher dust-to-gas ratio), but as we show here, an increase in the available solid material typically does not lead to further enrichment due to the low accretion rate. Protoplanets around stars with lower metallicity than considered here, will have a smaller fraction of heavy elements due to their initial composition which is metal-poor, and the fact that planetesimal accretion is inefficient as a result of the long accretion timescale and the low solid surface density. Our suggestion that massive gaseous planets at large radial distances will have stellar compositions is therefore independent of stellar metallicity.
 \par

Our conclusion that massive protoplanets ($\ge$ 3 M$_J$) will have relatively low 
enrichments in heavy elements is valid under the assumption that the high-Z
 material is captured during the pre-collapse stage in the form of solid
 planetesimals. It is possible however, that another enrichment mechanism for 
massive gaseous planets exists. For example, it may be possible that gaseous
 protoplanets are enriched with heavy elements from 'birth'. 
 Disk fragments form at the corotation of dense spiral waves (Durisen et al. 2008) which can be enhanced with 
 heavier elements (e.g., Haghighipour and Boss 2003, Rice et al.
2004) leading to formation of enriched fragments. In that case, solid concentrations are achieved at the locations where  fragmentation is most likely to occur, and the formed protoplanets will be metal-rich relatively to their host star. In addition, simulation of solids
 in disks show that regions in the disk can become highly enriched with solid
 material as a result of vortices in the disk (e.g., Klahr and Bodenheimer, 2006), although such vortices are likely to be destroyed in self-gravitating disks (Mamatsashvili and Rice, 2009). In any case, forming gaseous massive protoplanets which are enriched with heavy elements from birth is possible, and as a result massive gaseous protoplanets at large radial distances could contain a significant amount of heavy elements. 
It would be desirable to investigate whether enrichment from birth can be distinguished from enrichment at later stages (planetary evolution, internal structure, etc.), and we hope to address this issue in future work. An even more important issue that
needs to be addressed to make future progress on this problem
is a detailed study of planetesimal formation and collisional
evolution in the 50--100 AU region of a gravitationally unstable disk. 
 \par
 
Given the assumption that the planets around HR 8799 formed from near-stellar
composition, a question arises as to the observable metallicity in
the present atmospheres. Even if there is added planetesimal
material it tends to be deposited deep inside the planet, and
a substantial fraction of it would condense out and form a solid
core (Helled et al., 2008). Also, to a lesser extent, the grains
in the original composition of the planet would tend to
settle out, and at least the rock component could  be added to the core.
The heavy element material not in the core would have been mixed
by convection before the present time.
These considerations suggest a slight depletion in heavy elements
in the envelope of the planet as compared with the star. However
numerous additional processes, such as cloud formation and possible
core erosion, affect the atmosphere
between the early phase studied here and the present time. Thus it
is difficult to make precise predictions. In the limiting case where the
planet has been completely mixed and has uniform composition at
the present time, a typical case (38 AU, 10 km planetesimals,
7 M$_J$), the added solid mass is only 2.9 M$_\oplus$ compared to
a total mass of 2226 M$_\oplus$. 
Thus even with some enhancement
as discussed earlier, the difference between planetary and stellar
metallicity would be negligible. Even in the case of the most-enriched
planet, the fraction of solid material added is only 4\% by mass.

\section*{Acknowledgments}
R. H. acknowledges support from NASA through the Southwest Research Institute. P. B. acknowledges support from the NASA Origins grant NNX08AH82G.

\section*{References} 
\noindent Alexander, D., Ferguson, J., 1994. Low-temperature Rosseland opacities. 
Astrophys. J. 437, 879--891.\\
\noindent Bodenheimer, P., Grossman, A. S., Decampli, W. M., Marcy, G., Pollack,
 J. B., 1980. Calculations of the evolution of the giant planets. Icarus 41, 293--308.\\
Boley, A. C., 2009. The two modes of giant planet formation. Astrophys. J. 
695, L53--56. \\
Boss, A. P., 1997. Giant planet formation by gravitational instability. Science 276,
1836--1839.\\
Boss, A. P., 2002. Evolution of the solar nebula. V. Disk instabilities with varied
 thermodynamics.  Astrophys. J. 576, 462--472. \\
Cai, K., Pickett, M. K., Durisen, R. H., Milne, A. M., 2009.
 Giant planet formation by disk instability: A comparison simulation with an 
improved radiative scheme.  Astrophys. J. Letters, submitted. \\
Cameron, A. G. W., 1978. Physics of the primitive solar nebula and of giant
 gaseous protoplanets. In: Gehrels, T. (Ed.) Protostars and planets: Studies of star
 formation and of the origin of the solar system.  University of Arizona Press, 
Tucson, pp. 453--487. \\
Cuzzi, J. N., Hogan, R. C., Shariff, K., 2008. Toward planetesimals:
 Dense chondrule clumps in the protoplanetary nebula. Astrophys. J. 687, 1432--1447.\\
Decampli, W. M., Cameron, A. G. W., 1979. Structure and evolution of isolated giant
 gaseous protoplanets. Icarus 38, 367--391.\\
Dodson-Robinson, S. E., Willacy, K., Bodenheimer, P., Turner, N. J., Beichman, C. A., 2009(a). 
Ice lines, planetesimal composition and solid surface density in the solar nebula. Icarus, 200, 672--693.\\
Dodson-Robinson, S., Veras, D., Ford, E., Beichman, C. A., 2009(b). The formation
mechanism of gas giants on wide orbits. arXiv:0909.2662. \\
Durisen, R. H., Hartquist, T. W., Pickett, M. K., 2008. The formation of fragments at corotation in isothermal protoplanetary disks. 
Astrophysics and Space Science, 317, 3--8.\\
Greenzweig, Y., Lissauer, J.J., 1990. Accretion rates of protoplanets. Icarus 87, 40--77. \\
Guillot, T., 2007. The composition of transiting giant extra-solar planets. 
Physica Scripta 130, 014023--014029.\\
Haghighipour, N.,  Boss, A. P. 2003. On Gas Drag-Induced Rapid Migration of Solids in a Nonuniform Solar Nebula. 
ApJ, 598, 1301--1311.\\
Helled, R., Schubert, G., 2009. Heavy-element enrichment of a Jupiter-mass 
protoplanet as a function of orbital location.  Astrophys. J. 697, 1256--1262. \\
Helled, R., Podolak, M., Kovetz, A., 2006. Planetesimal capture in the disk 
instability model. Icarus 185, 64--71.\\
Helled, R., Podolak, M., Kovetz, A., 2008. Grain sedimentation in a giant gaseous
 protoplanet. Icarus 195, 863--870. \\
Henyey, L., Forbes, J., Gould, N., 1964. A new method of automatic computation of
stellar evolution. Astrophys. J. 139, 306--317.\\
Hubickyj, O., Bodenheimer, P., Lissauer, J. J., 2005. 
 Accretion of the gaseous envelope of Jupiter around a 5--10 Earth-mass core.
 Icarus 179,  415--431.\\
Ida, S., Lin, D. N. C. 2004. Toward a Deterministic Model of Planetary Formation. I. 
A Desert in the Mass and Semimajor Axis Distributions of Extrasolar Planets. ApJ, 604, 388--413.\\
Johansen, A., Oishi, J. S., Low, M.-M. M., Klahr, H., Henning, T., Youdin, A., 2007.
 Rapid planetesimal formation in turbulent circumstellar disks. Nature 448, 1022--1025.\\
Kalas, P. et al.,  2008.  Optical imaging of an exosolar planet 25 light 
years from Earth.  Science 322, 1345--47. \\
Klahr, H., Bodenheimer, P., 2006. Formation of giant planets by concurrent 
accretion of solids and gas inside an anticyclonic vortex. Astrophys. J. 639, 432--440.\\ 
Mamatsashvili, G. R., Rice, W. K. M., 2009. Vortices in self-gravitating gaseous discs. MNRAS, 394, 2153--2163.\\
Marley, M. S., Fortney, J. J., Hubickyj, O., Bodenheimer, P.,  Lissauer, J. J.,
2007. On the luminosity of young jupiters. Astrophys. J. 655, 541--549.\\
Marois, C., Macintosh, B., Barman, T.,  Zuckerman, B., Song, I., Patience, J.,
 Lafreniere, D., Doyon, R., 2008. Direct imaging of multiple planets orbiting the
 star HR 8799. Science 322, 1348--1352.\\
Mayer, L., Lufkin, G., Quinn, T.,   Wadsley, J., 2007. Fragmentation of 
gravitationally unstable gaseous protoplanetary disks with radiative transfer.
 Astrophys. J. 661, L77--L80.\\
Mayer, L., Quinn, T., Wadsley, J.,  Stadel, J., 2004. The evolution of 
gravitationally unstable protoplanetary disks: Fragmentation and 
possible giant planet formation.  Astrophys. J. 609, 1045--1064.\\
Mayer, L., Quinn, T., Wadsley, J.,  Stadel, J., 2002. Formation of giant planets
 by fragmentation of protoplanetary disks. Science 298, 1756--1759.\\
Morbidelli, A., Bottke, W. Nesvorny, D., Levison, H. F., 2009. Asteroids
were born big. Icarus, in press. \\
Nero, D., Bjorkman, J. E., 2009. Did Fomalhaut, HR8799, and HL Tauri form
planets via the gravitational instability?  arXiv:0908.1108.\\
Podolak, M., Pollack, J. B.,  Reynolds, R. T., 1987. The interaction of 
planetesimals with protoplanetary atmospheres. Icarus 73, 163--179.\\
Pollack, J. B., Hubickyj, O., Bodenheimer, P., Lissauer, J. J., Podolak, M.,
 Greenzweig, Y., 1996. Formation of the giant planets by concurrent accretion of
 solids and gas. Icarus 124, 62--85.\\
Pollack, J., McKay, C., Christofferson, B., 1985. A calculation of the Rosseland
mean opacity of dust grains in primordial Solar System nebulae.  Icarus 64, 
471-492.\\
Rafikov, R. R., 2007. Convective cooling and fragmentation of gravitationally 
unstable disks. Astrophys. J. 662, 642--650.\\ 
Rafikov, R. R., 2009. Properties of gravitoturbulent accretion disks.
 arXiv:0901.4739.\\
 Rice, W. K. M., Lodato, G., Pringle, J. E., Armitage, P. J., Bonnell, I. A., 2004. Accelerated planetesimal growth in self-gravitating protoplanetary discs. MNRAS, 355, 543--552. \\
Safronov, V. S., 1969. Evolution of the protoplanetary cloud and
formation of the Earth and planets. Nauka, Moscow. In Russian.
English translation: NASA-TTF-677, 1972.\\
Saumon, D., Chabrier, G., van Horn, H., 1995. An equation of state for low-mass
stars and giant planets. Astrophys. J. Suppl. 99, 713--741.\\

\clearpage

\begin{figure}[h!]
   \centering
    \includegraphics[width=7.0in]{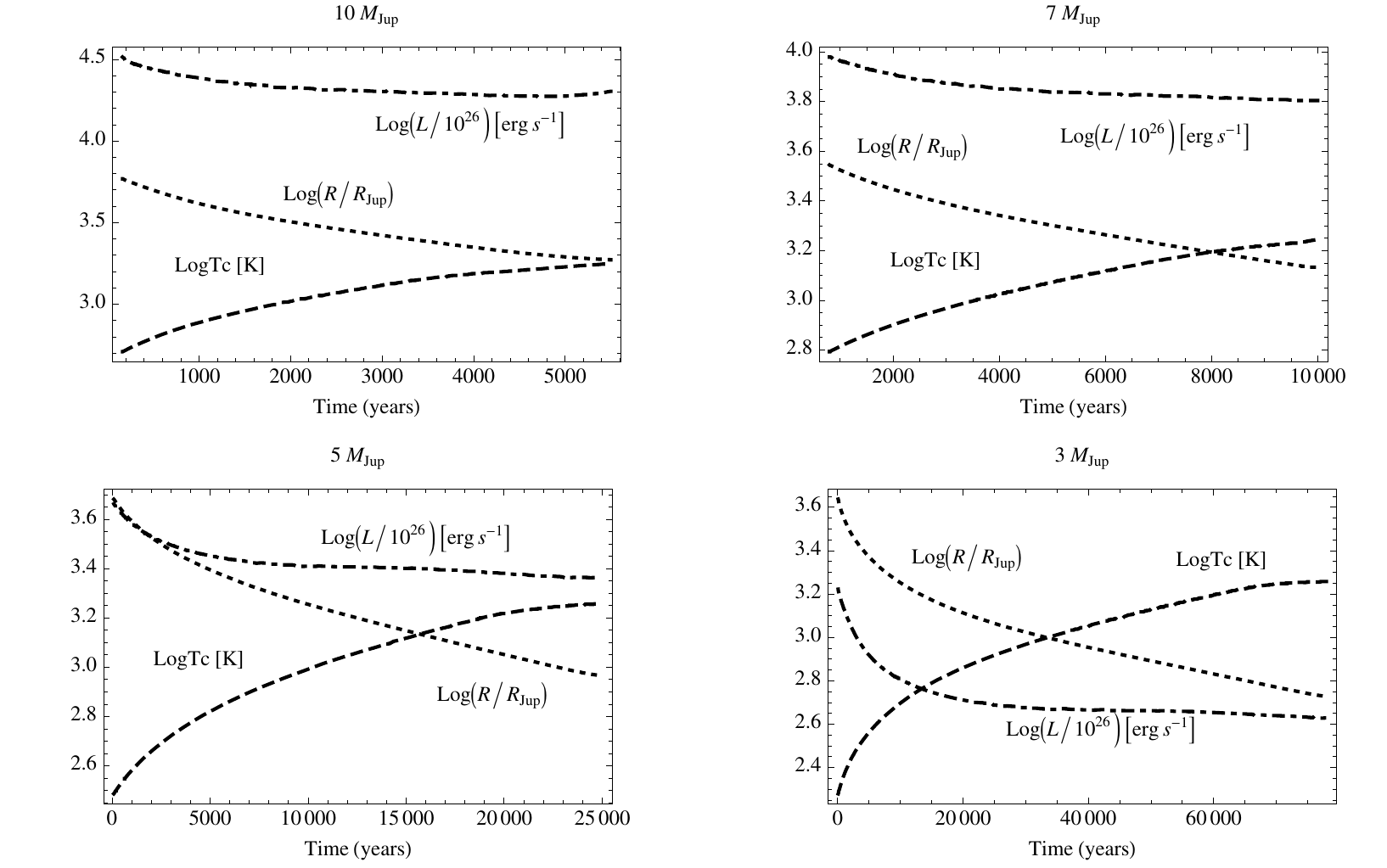}
    \caption[err]{Pre-collapse evolution of 3, 5, 7 and 10 Jupiter masses. The luminosity, radius, and central temperature as a function of time are presented.}
\end{figure}

\clearpage 
\clearpage
\begin{figure}[h!]
   \centering
    \includegraphics[width=5.0in]{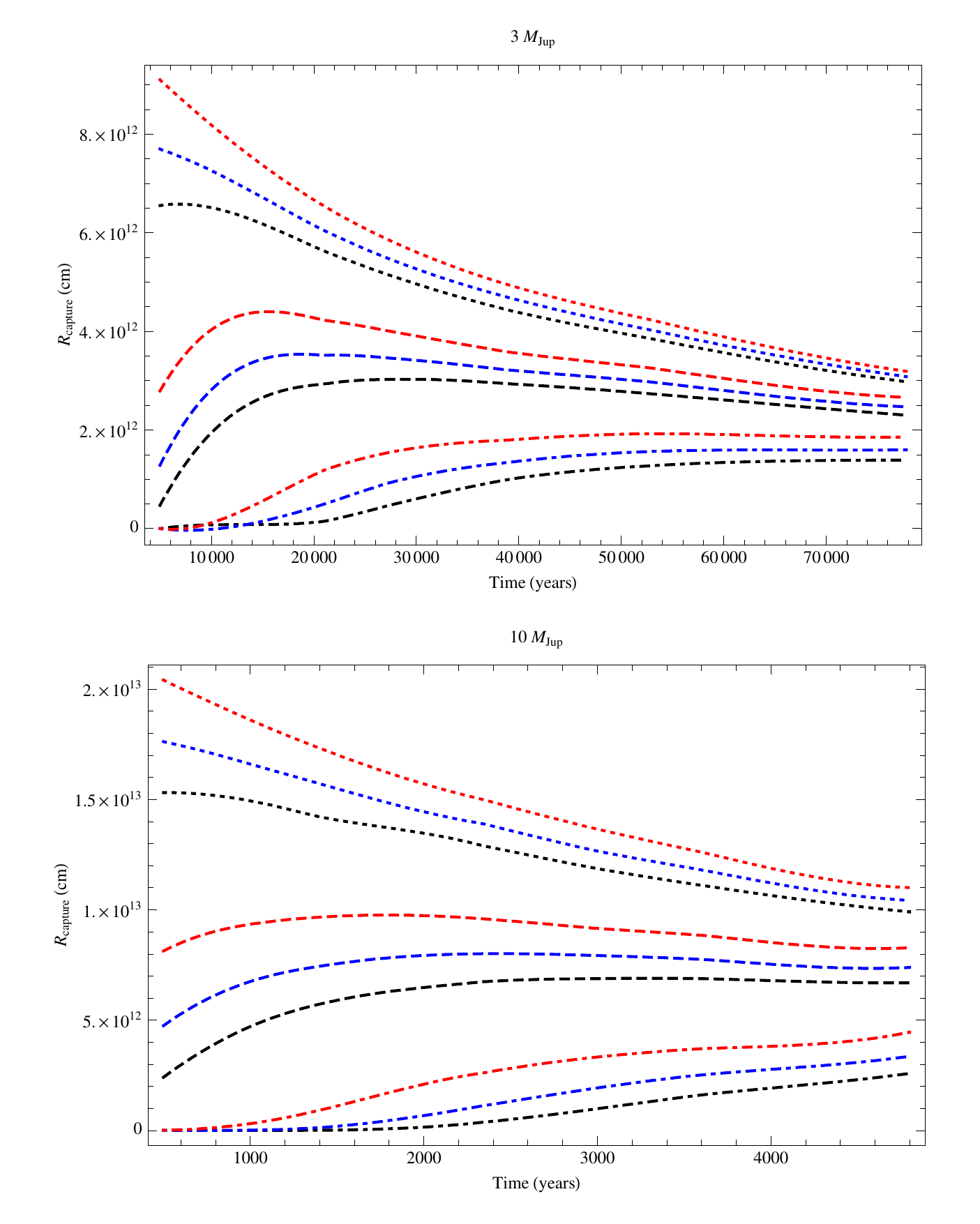}
    \caption[err]{Capture radius vs. time for 3 (top) and 10 (bottom) M$_J$ protoplanets.  The black, blue, and red curves correspond to radial distances of 24, 38, and 68 AU, respectively. The dotted, dashed, and dashed-dotted curves represent 1, 10, and 100 km-sized planetesimals, respectively.}
\end{figure}

\clearpage

\begin{table}[h!]
\centering
{\renewcommand{\arraystretch}{1.52}
\begin{tabular}{|c|c|c|c|c|}
\hline
Planetary Mass & Initial Radius & Central & Central Density & $\tau$ pre-collapse\\
(M$_{J}$) & ($\times 10^{13}$ cm) & Temperature (K) & ($\times 10^{-10}$ g cm$^{-3}$) & (years)  \\
\hline
3 & 3.16 & 188 & 3.83 & 7.82$\times10^{4}$\\
5 & 3.49 & 302 & 5.70 & 2.62$\times10^{4}$\\
5 - hot start & 0.83 & 1353 & 659.9 & 1.69$\times10^{4}$\\
7 & 3.23 & 499 & 14.0 & 1.17$\times10^{4}$\\
10 & 4.19 & 512 & 7.43 & 5.67 $\times10^{3}$\\
\hline
\end{tabular}
}
\caption{{\small Physical properties of the initial models for the four considered planetary masses. The fifth column gives the pre-collapse time-scale. }} \label{tab:1}
\end{table}

\clearpage

\begin{sidewaystable}[h!]
\small{
{\renewcommand{\arraystretch}{1.52}
\begin{tabular}{|c|c|c|c|c|c|c|c|c|c|c|c|}
\hline
r  & $\sigma$ & v$_{\inf}$$\times 10^4$ & planetesimal & M$_p$ &Captured & M$_p$ &Captured &M$_p$ &Captured & M$_p$ &Captured   \\
(AU) & (g cm$^{-2}$) & (cm s$^{-1}$) & size (km) & (M$_J) $& Mass (M$_{\oplus}$)& (M$_J) $& Mass (M$_{\oplus}$)& (M$_J) $& Mass (M$_{\oplus}$)& (M$_J) $& (M$_{\oplus}$) \\
\hline
24 &  7 &7.44 & 1& 10 &90.1 & 7 & 45.6 & 5 &  22.8 & 3 &  37.5\\
24 &  7 &7.44 & 10 & 10& 14.8 & 7& 8.7 & 5&  4.6 & 3&  5.0 \\
24 &  7 &7.44 & 100 & 10& 0.3 & 7& 0.8 & 5& 0.6 & 3&  1.0 \\
38 & 3 & 5.92 & 1 &  10& 24.7 & 7 & 12.9 & 5 &  7.0 & 3 &  10.6\\
38 & 3 & 5.92  & 10 & 10& 5.1 & 7 & 2.9 & 5 &  1.3 & 3 &  1.5\\
38 & 3 & 5.92 & 100 & 10 &  0.1& 7 & 0.3& 5 & 0.2 & 3 &  0.3\\
68 & 1 & 4.42 & 1 & 10& 4.3 & 7& 1.9 &5& 1.5& 3& 1.9\\
68 & 1 & 4.42 & 10 &10& 1.2 &7& 0.63&5& 0.2&3&  0.4\\
68 & 1 & 4.42 & 100 &10& 0.06 &7& 0.07 &5& 0.05 &3& 0.07\\
\hline
\end{tabular}
}
}
\caption{{\small Captured mass for 3, 5, 7, and 10 Jupiter masses protoplanets with 'cold start'. For comparison, the
initial complement of heavy elements, in the special case of
solar abundances, in a 5 M$_J$ planet is about 30 M$_{\oplus}$}} \label{tab:2}
\end{sidewaystable}

\clearpage

\begin{table}[h!]
\centering
{\renewcommand{\arraystretch}{1.52}
\begin{tabular}{|c|c|c|c|c|}
\hline
r  & $\sigma$ & v$_{\inf}$$\times 10^4$ & planetesimal &Captured   \\
(AU) & (g cm$^{-2}$) & (cm s$^{-1}$) & size (km) & Mass (M$_{\oplus}$)\\
\hline
24 &  7 &7.44 & 1& 18.4\\
24 &  7 &7.44 & 10 & 9.2\\
24 &  7 &7.44 & 100 & 1.6\\
38 & 3 & 5.92 & 1 &  4.3\\
38 & 3 & 5.92  & 10 & 2.3\\
38 & 3 & 5.92 & 100 & 0.6\\
68 & 1 & 4.42 & 1 & 0.7\\
68 & 1 & 4.42 & 10 & 0.4\\
68 & 1 & 4.42 & 100 & 0.1\\
\hline
\end{tabular}
}
\caption{{\small Test case: captured mass for 5 Jupiter masses protoplanets 
with 'hot start'.}} \label{tab:3}
\end{table}

\end{document}